# Performance Comparison of Reed Solomon Code and BCH Code over Rayleigh Fading Channel

Faisal Rasheed Lone
M.Tech.scholar Department of Computer Science and Engineering, Shri Mata Vaishno Devi University, Katra, Jammu and Kashmir, India

Arjun Puri
M.tech. scholar Department of Computer Science and Engineering, Shri Mata Vaishno Devi University, Katra, Jammu and Kashmir, India

Sudesh Kumar
Lecturer Department of Computer Science and Engineering, Shri Mata Vaishno Devi University,Katra, Jammu and Kashmir, India

## ABSTRACT
Data transmission over a communication channel is prone to a number of factors that can render the data unreliable or inconsistent by introducing noise, crosstalk or various other disturbances. A mechanism has to be in place that detects these anomalies in the received data and corrects it to get the data back as it was meant to be sent by the sender. Over the years a number of error detection and correction methodologies have been devised to send and receive the data in a consistent and correct form. The best of these methodologies ensure that the data is received correctly by the receiver in minimum number of retransmissions. In this paper performance of Reed Solomon Code (RS) and BCH Code is compared over Rayleigh fading channel.

## General Terms
Rayleigh Fading Model for RS and BCH Code.

## 1. INTRODUCTION
Whenever communication takes place, data needs to be transmitted, be it human beings or computers. Data is of utmost importance for an effective communication to take place, so the transmission of data should be such that the receiver of the data should receive the data in the same condition as it was sent by the sender. If the data is prone to any disturbance, then the data received by the receiver will not be the same as it was meant and thus it will convey a different information than what it was meant to convey. Thus the goal of data transmission is to transmit the data over a communication channel without any errors. Various techniques have been developed over the past few years to secure and make the data transmission reliable. One of such techniques is Information Coding Theory. Coding theory the study of codes, including error detecting and error correcting codes, has been studied extensively for the past forty years. It has become increasingly important with the development of new technologies for data communications and data storage. Coding theory makes use of various codes to encode the data for transmission over a channel and then the data is decoded at the receivers end to get the required data bits. These codes used to encode the data bits show variations over different communication channel's, that is the behavior of these codes may be different in different communication channels. The performance of some codes may be better than other codes over the same channel. This paper discusses the performance comparison of BCH code and Reed Solomon code over Rayleigh fading channel.

## 2. RELATED WORK
In the recent years, researchers have shown their keen interest towards analyzing the performance of various forward error correction techniques .They used various communication channels to analyze the behavior of error correcting codes.

In [5] authors have considered various concatenated error correcting codes using Binary Phase Shift Keying (BPSK) modulation scheme like Convolutional-Hamming, Convolutional-Cyclic, Convolutional-Bose chaudhuri Hocquenghem was designed and the BER performance was measured for an Additive White Gaussian Noise (AWGN) channel. In general Convolutional-Bose Chaudhuri Hocquenghem demonstrated better performance compared to Convolutional-Hamming and Convolution-Cyclic concatenation pair

In [6] paper authors performed Reed-Solomon code for M-ary modulation in a communication noisy channel (AWGN). In this authors performed simulation on MATLAB and concluded following results:-

(a) The BER performance curve improves as the code size (n) increases at constant code rate and at the similar error correcting capability (t).

(b) The BER performance curve improves as the redundancy of the codes increases.

(c) The BER performance curve improves as the code rate decreases.

In [7], authors studied performance of RS code and calculated the Symbol Error Rate (SER), the performance was evaluated and compared with theoretical calculations. In this they concluded that simulated RS code SER shows good performance over AWGN channel as compared to Rayleigh Channel.

## 3. REED SOLOMON CODE

lrving S. reed and Gustave Solomon in 1960[4] developed a code for data transmission known as Reed Solomon Code. Block coding forms the basis of this code. A block of k information bits is taken and then encoded into a block of n bits known as codeword (n>k) which is used for actual transmission. As there are k bits in the block $2^k$ codeword's are possible. RS codes find their use in digital communication and storage. This code finds its application in space communication, storage media, wireless communication, digital television etc. RS code is demonstrated below:-

RS (n, k) codes on m-bit symbols exist for all n and k for which:-

$$0<k<n<2^m + 2 \quad (1)$$

Here $k$ is the number of message bits to be encoded, $n$ is the size of code word in an encoded block and m is the number of bits per symbol. Thus RS (n,k) can be written as follows:-

$$(n, k) = (2^{(m-1)}, 2^{(m-1)} - 1 - 2t) \quad (2)$$

Number of parity bits added to the message bits is calculated by (n-k) = 2t where t is the number of errors corrected by RS code.

The distance of the RS code is given by:-

$$d_{min} = n-k+1 \quad (3)$$

thus the minimum distance of RS code is similar to hamming distance. Reed-Solomon code is based on galoi's field, According to galoi's a finite field has the property that arithmetic operations (+,-,*, / etc.) on the field elements always have a result within the field itself.

### 3.1 RS Encoder

If a finite field of q elements is chosen, whose $GF(2^m)$, as a result the message f to be transmitted, consists of k elements of $GF(2^m)$ which are given by:-

$$f = (f_0, f_1, \ldots f_{k-1}) \quad (4)$$

Where $f_i \in GF(2^m)$

Thus message polynomial is calculated by multiplying coefficients of the message with appropriate power of x which is given as follows:-

$$F(x) = f_0 + f_1 x + \_\_\_\_\_\_ + f_{k-1} x^{k-1} \quad (5)$$

The remaining polynomial is known as parity check polynomial:-

$$B(x) = b_0 + b_1 x + \_\_\_\_\_\_ + b_{2t-1} x^{2t-1} \quad (6)$$

Then the codeword is form by adding the two polynomials as follows:-

$$V(x) = F(x) + B(x) \quad (7)$$

### 3.2 RS Decoder

When the message is being transmitted a lot of issues can arise such as corruption of the sent message due to a noisy channel etc..Thus the received message at the receivers end is r(x) which is given by following expression:-

$$R(x) = C(x) + E(x) \quad (8)$$

C(x) corresponds to the original codeword transmitted and E(x) is the error, which is further given by expression given as follows:-

$$E(x) = e_{n-1} x^{n-1} + \_\_\_\_\_\_\_\_ + e_1 x + e_0 \quad (9)$$

Using RS Code t= {n-k}/2 errors can be corrected, if errors are more than t then this code fails [6].

## 4. BCH CODE

Bose – Chaudhuri – Hocquenghem [8] developed BCH Code. Multiple errors can be detected and corrected using BCH code.BCH Code is a generalized form of Hamming Code. The possible BCH codes for m>=3 and t<$2^{m-1}$ are:-

Block length: $n=2^m-1$
Parity check bits: $n-k<=mt$
Minimum distance: $d>= 2t+1$

Generating polynomial g(x) is generally created as follows:-

$$L.C.M \text{ of } \{m_1(x), m_2(x) \ldots m_{2t-1}(x)\}$$

The message m(x) is divided by g(x) and remainder will be represented as check bits r(x). Now whole encoded message E(x) will be represented as:-

$$E(x) = m(x) + r(x)$$

### 4.1 BCH Decoding

The Decoding of BCH code is performed in three steps:-
* Syndrome is calculated from the received codeword.
* Error location polynomial is found from a set of equation derived from the syndrome.
* Error location polynomial is used to identify and correct the errant bits[1].

## 5. METHODLOGY

The tool used for performing simulation of coding and decoding of Reed-Solomon and BCH codes through BPSK and QAM modulation scheme in Rayleigh channel was SIMULINK in MATLAB. The process is described as follows:

K random information symbols are input to the encoder for transmission. The encoder maps each of the input sequences to unique n symbol sequence known as a Codeword. The generated Codeword is then passed to the next module known as a digital modulator. The modulator uses modulation schemes like BPSK and QAM to modulate the data into signal waveforms. After the data has been transformed to signals it is sent over the Rayleigh fading channel for transmission The channel is prone to various noises like man made noises and other disturbances which can corrupt or alter the data Before the sent data can be decoded it had to be separated from the carries waves using a demodulator.. After demodulation is performed and the data separated from the carrier waves, its time for decoding the sent data. The demodulated data is then sent to the decoder at the receiver's end which decodes data into original information sequence. This also depends upon system that we are going to use e.g. size of block etc. In this simulation bit error ratio is calculated.

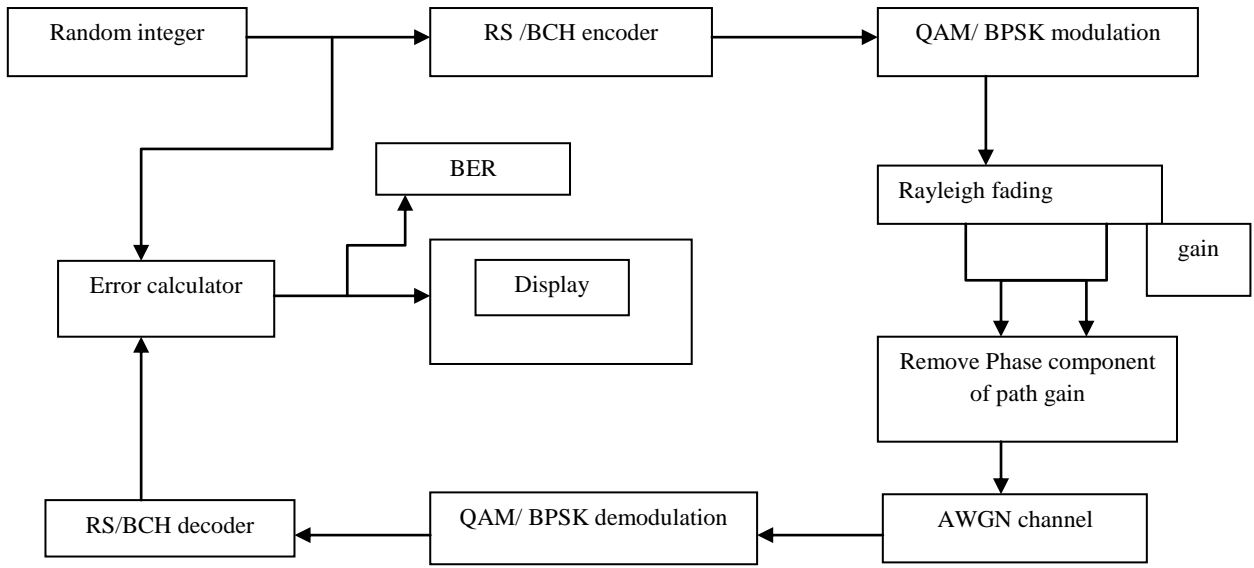

**Figure 1:-Proposed Simulation Model**

## 6. RESULT AND DISCUSSION

In this paper analysis on RS and BCH code of block length (15, 11) was performed. BER ratio was calculated by varying $E_b/N_0$ from 0 to 10. Doppler shift value was set as .0001. The analysis results are given in table 1:-

**TABLE 1:- (left) Performance Comparison for BCH and RS codes with BPSK modulation and (right) ) Performance Comparison for BCH and RS codes with QAM modulation in the presence of Rayleigh Fading channel.**

| $E_b/N_0$ | RS_QAM | BCH_QAM | $E_b/N_0$ | RS_BPSK | BCH_BPSK |
|---|---|---|---|---|---|
| 0 | 0.8802 | 0.2835 | 0 | 0.8975 | 0.2823 |
| 1 | 0.8719 | 0.2235 | 1 | 0.8906 | 0.2273 |
| 2 | 0.8727 | 0.09173 | 2 | 0.8922 | 0.09306 |
| 3 | 0.8636 | 0.08569 | 3 | 0.8826 | 0.08601 |
| 4 | 0.8556 | 0.05811 | 4 | 0.8727 | 0.05854 |
| 5 | 0.8443 | 0.03972 | 5 | 0.8619 | 0.04063 |
| 6 | 0.8289 | 0.03739 | 6 | 0.846 | 0.03721 |
| 7 | 0.8061 | 0.03548 | 7 | 0.8392 | 0.03484 |
| 8 | 0.7913 | 0.02272 | 8 | 0.8239 | 0.02296 |
| 9 | 0.7621 | 0.01703 | 9 | 0.7913 | 0.01706 |
| 10 | 0.7215 | 0.01677 | 10 | 0.7513 | 0.01686 |

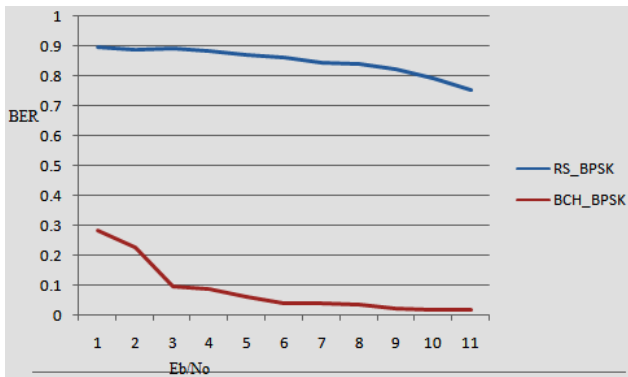 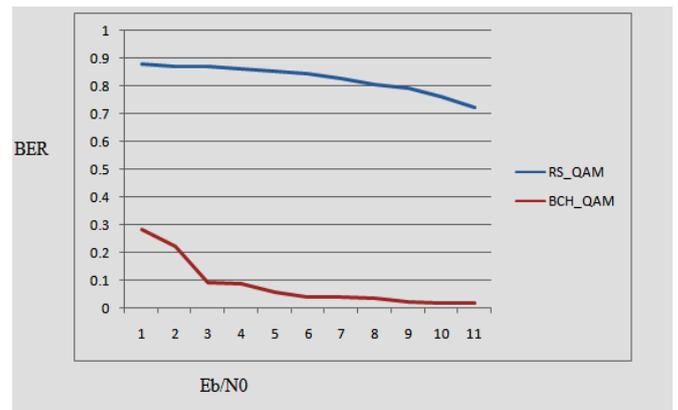

**Figure 2 :( left) BER of RS and BCH (15, 11) code with QAM modulation and (Right) BER of RS and BCH (15, 11) code with BPSK modulation in Rayleigh Fading channel.**

From the above graph it was found that BCH code consistently performed better than RS Code in presence of Rayleigh channel under binary environment.

## 7. CONCLUSION

In this paper performance of RS and BCH code was simulated in the presence of Rayleigh fading channel. Several iterations were performed to find out that BCH code outperforms RS code in binary environment. Eb/No ratio was varied from 1 to 10 and it was noticed that at every value of $E_b/N_o$ BCH code performed considerably better than RS code. Both BPSK and QAM modulation schemes were used and BCH code again performed better in both modulation schemes. As $E_b/N_o$ was increased beyond 1 performance of BCH code improved drastically while there was only slight improvement in the performance of RS code. The graph plotted between RS code and BCH code shows the performance gain of BCH code over RS code in Rayleigh fading channel under BPSK and QAM modulation.

## 8. REFERENCES

[1] Hank Wallace, "Error detecting and correcting using BCH codes", Copyright (C) 2001 Hank Wallace.

[2] Sanjeev Kumar, Ragini Gupta,"Performance Comparison of Different Forward Error Correction Coding Techniques for Wireless Communication Systems", at International Journal of Computer science and technology Vol. 2, issue3, September 2011.

[3] Claude E. Shannon (1948) "A Mathematical Theory of communication", at bell system journal, vol.24, July and October 1948.

[4] I.S. Reed; G.Solomon, "polynomial codes over certain finite fields", Journal of the society for industrial and applied Mathematics, Vol. 8, No. 2 (jun., 1960), 300-304.

[5] Himanshu Saraswat, Govind Sharma, Sudhir Kumar Mishra and Vishwajeet, "Performance Evaluation and Comparative Analysis of Various Concatenated Error Correcting Codes Using BPSK Modulation for AWGN Channel", International Journal of Electronics and Communication Engineering. ISSN 0974-2166 Volume 5, Number 3 (2012), pp. 235-244.

[6] Saurabh Mahajan; Gurpadam Singh "Reed-Solomon Code Performance for M-ary Modulation over AWGN Channel", International Journal of Engineering Science and Technology (IJEST), Vol. 3 No. 5 May 2011.

[7] Gurinder Kaur Sodhi ; Kamal Kant Sharma, "SER performance of Reed – Solomon Codes With AWGN & Rayleigh Channel using 16 QAM" ,at International Journal of Information and Telecommunication Technology, Vol. 3, No. 2, 2011.

[8] R.C. Bose and D.K. Ray-Chaudhuri, " On A Class of Error Correcting Binary Group Codes", at Information and Control 3, 68-79 (1960).

[9] Lionel Biard, Dominique Noguet, "Reed-Solomon Codes for Low Power Communications", Journal of communications, vol. 3, no. 2, April 2008

[10] Arjun Puri, Sudesh Kumar, "Comparative Analysis of Reed Solomon Codes and BCH Codes in the Presence of AWGN Channel", International Journal of Information and Computing Technology, vol.3 no.3, 2013

[11] Zhinian Luo, Wenjun Zhang,"The Simulation Models for Rayleigh Fading Channels" IEEE Transactions on Communications, Vol. 61, No. 2, February 2013

[12] Shih-Ming Yang and Vinay Anant Vaishampayan "Low-Delay Communication for Rayleigh Fading Channels: An Application of the Multiple Description Quantizer"

[13] J Grolleau, D. Labarre, E. Grivel and M. Najim,"The stochastic sinusoidal model for Rayleigh fading channel simulation"